\documentclass[a4paper,12pt]{article}
\usepackage[reqno]{amsmath}
\usepackage{mathrsfs}
\usepackage{amssymb}
\usepackage{url}

\newcommand{\Rep}[1]{\underline{\mbox{\textbf{#1}}}}

\usepackage{amsmath}
\usepackage{amssymb}
\numberwithin{equation}{section}

\usepackage{epsf}
\usepackage{color}
\usepackage{graphicx}
\usepackage{amsmath}
\usepackage{amssymb}
\usepackage{latexsym}

\begin{document}

\setlength{\baselineskip}{18pt}
\begin{titlepage}

\vspace*{1.2cm}
\begin{center}
{\Large\bf {\boldmath $A_4$} Flavor Models in Split Seesaw Mechanism}
\end{center}
\lineskip .75em
\vskip 1.5cm

\begin{center}
{\large
Adisorn Adulpravitchai\footnote[1]{E-mail:
\tt adisorn.adulpravitchai@mpi-hd.mpg.de}
and
Ryo Takahashi\footnote[2]{E-mail:
\tt ryo.takahashi@mpi-hd.mpg.de }
}\\

\vspace{1cm}

{\it Max-Planck-Institut f$\ddot{u}$r Kernphysik, Postfach 10 39 80,
69029 Heidelberg, Germany}

\vspace*{10mm}
{\bf Abstract}\\[5mm]
{\parbox{13cm}{\hspace{5mm}
A seesaw mechanism in an extra-dimension, known as the split seesaw mechanism, 
provides a natural way to realize a splitting mass spectrum of right-handed
 neutrinos. It leads to one keV sterile neutrino as a dark matter candidate and 
two heavy right-handed neutrinos being responsible for leptogenesis to explain 
the observed baryon asymmetry of the Universe. We study models based on $A_4$ 
flavor symmetry in the context of the split seesaw mechanism. It is pointed out 
that most of known $A_4$ flavor models with three right-handed neutrinos being 
$A_4$ triplet suffer from a {\it degeneracy problem} for the bulk mass terms, 
which disturbs the split mechanism for right-handed neutrino mass spectrum. Then
 we construct a new $A_4$ flavor model to work in the split seesaw mechanism. In
 the model, the experimentally observed neutrino masses and mixing angles can be
 realized from both type I+II seesaw contributions. The model predicts the 
$\mu-\tau$ symmetry in the neutrino mass matrix at the leading order, resulting 
in the vanishing  $\theta_{13}$ and maximal $\theta_{23}$. The flavor symmetry 
$A_4$ is broken via the flavon vacuum alignment which can be obtained from the 
orbifold compactification. The model can be consistent with all data of neutrino
 oscillation experiments, cosmological discussions of dark matter abundance, 
leptogenesis, and recent astrophysical data.
}}
\end{center}
\end{titlepage}
\newpage

\section{Introduction}

Extra-dimensional theory is a fascinating approach that affords a compelling 
candidate solution to the hierarchy problem~\cite{ArkaniHamed:1998rs}. Indeed, 
the discovery of an evidence of extra-dimension as well as the Higgs particle 
and supersymmetry (SUSY) is one of the important missions of the CERN Large 
Hadron Collider (LHC) experiment. In the extra-dimensional theory, there are 
also some alternatives to the ordinary electroweak symmetry breaking (EWSB) 
mechanism in the standard model (SM), such as the gauge-Higgs unification 
(GHU)~\cite{Fairlie:1979at}, the little Higgs~\cite{ArkaniHamed:2001nc}, the 
Higgsless~\cite{Csaki:2003dt}, and the Dirichlet Higgs~\cite{Haba:2009pb} 
models, and so on. Furthermore, extra-dimensional models can give 
phenomenologically interesting features and predictions for 
physics beyond the SM, for example, the presences of 
Kaluza-Klein particles and a candidate for dark matter 
(DM)~\cite{Zwicky:1933gu} from the Universal Extra-Dimensions (UED) 
model~\cite{Antoniadis:1990ew}, and deviations of couplings of the 
Higgs~\cite{Haba:2009pb,Hosotani:2008tx}.

The elucidation of the origin of DM, which governs about 23\% of the 
Universe~\cite{Komatsu:2010fb}, is one of important goals of the particle 
physics today. A large number of DM candidates have been discussed in 
literature.\footnote{See ref.~\cite{Garrett:2010hd}, for a recent review, and 
references therein.} One of the interesting candidates for DM is a keV sterile 
neutrino since some astrophysical data possibly support the 
existence~\cite{Kusenko:2009up,Kusenko}. In extentions of the SM, the 
right-handed neutrinos can be added. The models with three right-handed 
neutrinos whose masses are below the EW scale have been proposed 
in~\cite{deGouvea:2005er,Asaka:2005an,Asaka:2005pn}. In particular, the
 ref.~\cite{Asaka:2005pn} has pointed out that the model with one keV sterile 
(right-handed) neutrino and two GeV range of ones can explain simultaneously the
 DM and baryon asymmetry of the Universe (BAU) in addition to the results of 
neutrino oscillation experiments. This framework is known as 
$\nu$MSM \cite{Asaka:2005an,Asaka:2005pn}.
Furthermore, the keV sterile neutrino has been also considered in a gauge 
extension of the SM (left-right symmetric framework) \cite{Bezrukov:2009th}. A 
realization of splitting pattern of right-handed neutrino mass spectrum 
including keV mass state in the context of flavor symmetry has been also 
discussed in \cite{Shaposhnikov:2006nn,Lindner:2010wr,Merle:2011yv}. In 
\cite{Shaposhnikov:2006nn}, it has been shown that the degeneracy of two 
singlet Majorana neutrinos and the lightness (keV scale) of the third one in 
$\nu$MSM can be a consequence of a lepton number symmetry, which is broken in 
both the Yukawa and Majorana mass sectors. In \cite{Lindner:2010wr}, a keV 
sterile neutrino as a candidate for DM is induced from softly breaking of 
$L_e-L_\mu-L_\tau$ flavor symmetry. In \cite{Merle:2011yv}, a realization of 
the keV sterile neutrino arises from the Froggatt-Nielsen mechanism 
\cite{Froggatt:1978nt}. Other interesting direction to realize the keV sterile 
neutrino DM and BAU has been discussed in the split seesaw mechanism 
\cite{Kusenko:2010ik}. This mechanism can realize a splitting mass spectrum of 
the right-handed neutrinos in the context of extra-dimension and account for 
the smallness of active neutrino masses without fine-tuning. In the mechanism, 
the lightest right-handed neutrino with keV mass becomes DM candidate and the 
other two right-handed neutrinos make a source of BAU via leptogenesis 
\cite{Fukugita:1986hr}. 

Regarding with properties of active neutrinos, various neutrino oscillation 
experiments have shown that the neutrino mixing pattern is peculiar, namely, 
there are two large mixing angles and one small one~\cite{Schwetz:2011qt}. A lot
 of works have been proposed to explain such a peculiar mixing pattern, e.g. 
$\mu-\tau$ symmetric texture~\cite{Fukuyama:1997ky}, tri-bimaximal 
mixing~\cite{Harrison:2002er}, golden ratios~\cite{goldenratio}, and etc.. A 
proposed mixing pattern can be explained by imposing a non-Abelian discrete 
flavor symmetry with its breaking in a specific direction. For example, 
$\mu-\tau$ symmetry can be explained by the group $S_3 \simeq D_3$~\cite{D3} or
 $D_4$~\cite{D4}. The tri-bimaximal mixing can be obtained from $A_4$~\cite{A4} 
or $S_4$~\cite{Ma:2005pd}. The golden ratios have been discussed with 
$A_5$~\cite{A5} and $D_{10}$~\cite{D10}.\footnote{Our references for flavor 
models based on non-Abelian discrete symmetry are not complete. See 
e.g.~\cite{Altarelli:2010gt} for more complete list.} Such the non-Abelian 
discrete symmetries might arise from the breaking of the gauge 
symmetries~\cite{ConSym} or from the orbifold compactification in 
extra-dimensions~\cite{OrbifoldSym}. Moreover, the vacuum alignment of the 
flavon field can be achieved without dealing with a relatively complicated 
scalar potential for $A_4$~\cite{A4breaking} and 
$S_4$~\cite{Adulpravitchai:2010na} symmetries in the extra-dimensional theory.

In this paper, we study models based on a flavor symmetry in the context of the 
split seesaw mechanism which can lead to a splitting mass spectrum of the 
right-handed neutrinos.\footnote{The ref.~\cite{Barry:2011wb} 
has studied the realization of eV scale sterile neutrinos within both seesaw 
mechanism and flavor symmetry. The work has shown that light sterile neutrinos 
can be accommodated in $A_4$ flavor models.} The paper is organized as 
following: In section 2, a brief review of the split seesaw mechanism is given. 
In section 3, we discuss the $A_4$ flavor models in the split seesaw mechanism. 
Then it will be shown how the desired flavon vacuum alignment can be obtained. 
We also give a comment on the leptogenesis. Section 4 is devoted to the 
conclusion.
  
\section{Split Seesaw Mechanism}
In this section, we give a brief review of the split seesaw 
mechanism~\cite{Kusenko:2010ik}. The relevant terms in the Lagrangian for the 
canonical type I seesaw mechanism~\cite{seesaw} reads
\begin{equation}
 {\cal L} \;=\; 
i {\bar N}_i  \gamma^\mu \partial_\mu N_i+
\left(
 \lambda_{i \alpha} {\bar N}_i L_\alpha \, \phi 
-\frac{1}{2} M_{R,ij} {\bar {N^c_i}} N_j + {\rm h.c.}
\right),
\end{equation}
where $N_i$ ($i$=1,2,3), $L_{\alpha}$ ($\alpha=e,\mu,\tau$), and $\phi$ are the 
right-handed neutrinos, lepton doublets, and the SM Higgs doublet, respectively.
 The $i$ corresponds to the generation of the right-neutrinos, and here we 
introduce three right-handed neutrinos to the SM. After integrating out the 
heavy right-handed neutrinos, the light neutrino mass matrix is given by
\begin{equation}
  M_{\nu}=\lambda^TM_R^{-1}\lambda\langle\phi^0\rangle^2,
\end{equation}
where the neutrino masses depend on the Yukawa coupling $\lambda$, the 
right-handed neutrino mass scale, and the vacuum expectation value (VEV) of the 
SM Higgs field. The seesaw mechanism can lead to small active neutrino mass 
with the order one Yukawa coupling and heavy right-handed neutrino mass. 
Moreover, the decay of the right-handed neutrino can also explain the BAU, if 
the mass of the right-handed neutrino is at $\mathcal{O}(10^{11-12})$ GeV 
\cite{Endoh:2002wm}. This process is well known as leptogenesis.  

Now, let us move on to an extra-dimensional theory compactified on the orbifold 
$S^1/(Z_2 \times Z_2')$ with an extra-dimensional coordinate, $y \equiv x^5$. 
Note that the original model of split seesaw mechanism~\cite{Kusenko:2010ik} 
has considered $S^1/Z_2$. An additional $Z_2'$ in our case is imposed in order 
to break the flavor symmetry as we will discuss in section~\ref{FlavorOrbifold}.
 However, the introduction of the additional $Z_2'$ does not change an essential
 point of the splitting mechanism. The circle $S^1$ has a radius $R$. The 
orbifold $S^1/Z_2$ is obtained by modding out the $Z_2$ transformation, which 
imposes the equivalence relation, $y\sim-y$. The orbifold $S^1/(Z_2\times Z_2')$
 is obtained by modding out further the $Z_2'$ transformation. It imposes the 
equivalence relation, $y'\sim-y'$, where $y'\equiv y+\pi R/2$. The fundamental 
region of this orbifold is given by $y\in[0,\ell\equiv\pi R/2]$. The size of the
 extra-dimension $\ell$ and the five dimensional fundamental scale $M$ are 
related to the 4D reduced Planck scale as $M_{pl}^2=M^3 \ell$. The orbifold gives
 two fixed points (branes). One of the branes $(y=0)$ is the SM brane where the 
SM particles reside, while the other $(y=\ell)$ is the hidden brane. 

Next, a Dirac spinor, $\Psi(y,x)=(\chi_{\alpha},\bar{\psi}^{\dot{\alpha}})^T$, is 
introduced in the bulk with a bulk mass term $m$,
 \begin{equation}
  S=\int d^4xdyM(i\bar{\Psi}\Gamma^A\partial_A\Psi+m\bar{\Psi}\Psi), 
  \label{action} 
 \end{equation}
where $A=0,1,2,3,5$, and the five-dimensional gamma matrices $\Gamma^A$ are 
defined as following
\begin{equation}
\Gamma^\mu = \begin{pmatrix} 0 & \sigma^{\mu} \\ \bar{\sigma}^{\mu} & 0 \end{pmatrix},\;\; \Gamma^5 =-i \begin{pmatrix} 1 & 0 \\ 0 & -1 \end{pmatrix}.
\end{equation}
Note that the mass scale $M$ in \eqref{action} is inserted so that the mass 
dimension of $\Psi$ is $3/2$ as in the 4D case. The zero mode of $\Psi$ follows 
the Dirac equation as
\begin{equation}
(i\Gamma^5 \partial_5 + m )\Psi^{(0)}=0,
\end{equation}  
leading to the wavefunction profile of the zero mode in the bulk as 
$e^{\mp my}$
for $\chi$ and $\bar{\psi}$. In order to get the chiral fermion in 4D, we 
promote that the field is transformed under the orbifold parities as
 \begin{eqnarray}
  Z_2:~\Psi\rightarrow P\Psi=+\Psi,~~~
  Z_2':~\Psi\rightarrow P'\Psi=+\Psi,
\end{eqnarray}
where $P = -i \Gamma_5$ and $P' = 1$. As the results, only $\bar{\psi}$ can have
 a zero mode. Note that the bulk mass term $m$ also has to carry negative $Z_2$ 
parity, so that the action is invariant under the above parity transformations.
 The bulk mass term having the negative parity can be realized from the kink 
profile of scalar field along the extra-dimensional direction. After canonically
 normalizing the fermion in 4D, the zero mode of $\Psi_R=(0,\bar{\psi})^T$ can 
be written in term of canonically normalized field $\psi_R^{(4D)}(x)$ as
 \begin{equation}
  \Psi_R^{(0)}(y,x)=\sqrt{\frac{2m}{e^{2m\ell}-1}}\frac{1}{\sqrt{M}}
                   e^{m y} \psi_R^{(4D)}(x), \label{5Dto4DSpinor}
 \end{equation}
where $\psi_R^{(4D)}(x)$ is identified with the right-handed neutrino in 4D. The 
point is that the extra-dimensional wavefunction profile of $\Psi_R^{(0)}(y,x)$ 
with real positive $m$ localizes on the hidden brane since the profile is 
expressed by the exponential function, $e^{my}$. Then $\psi_R^{(4D)}(x)$ couples 
to particles in the SM brane through exponentially suppressed couplings when 
$m\ell\gg1$.

Now let us show the action for the three right-handed neutrinos,
 \begin{eqnarray}
  S &=& \int d^4x\,dy\left\{M
        \left(i\bar{\Psi}^{(0)}_{iR}\Gamma^A\partial_A\Psi_{jR}^{(0)} 
              +m_{i}\bar{\Psi}_{iR}^{(0)}\Psi_{iR}^{(0)}\right)
        \right.\nonumber\\ 
    & & \phantom{\int d^4x\,dy\Big\{}\left.
        +\delta(y)\left(\frac{\kappa_{ij}}{2}v_{\rm B-L}\bar{\Psi}^{(0)c}_{iR}
                        \Psi^{(0)}_{jR}
                        +\tilde{\lambda}_{i\alpha}\bar{\Psi}^{(0)}_{iR}L_\alpha\phi 
                        +{\rm h.c.}\right)\right\}, \label{Sneutrino}
 \end{eqnarray}
where we work in the basis of the diagonal bulk mass term $m_i$. The Majorana 
mass of the right-handed neutrino arises from the breaking of $U(1)_{\rm B-L}$ 
symmetry at a high scale. Note that $v_{\rm B-L}$ determines the $B-L$ breaking
 scale. Inserting the result of \eqref{5Dto4DSpinor} into \eqref{Sneutrino}, 
the effective 4D mass and Yukawa coupling matrices are obtained 
\cite{Hebecker:2001wq},
 \begin{eqnarray}
  M_{R,ij}         &=& \kappa_{ij}f_if_jv_{\rm B-L}, \label{MR4D} \\
  \lambda_{i\alpha} &=& \tilde{\lambda}_{i\alpha}f_i, \label{Yukawa4D}
 \end{eqnarray}
where 
 \begin{eqnarray}
  f_i\equiv\frac{1}{\sqrt{M}} \sqrt{\frac{2 m_i}{e^{2 m_i \ell}-1}}.
 \end{eqnarray}
For the sake of the following discussion, we define a diagonal matrix $F_{ij}$ as
 \begin{equation}
  F_{ij}\equiv\delta_{ij}f_j.
 \end{equation}
The effective 4D mass and Yukawa coupling matrices given in \eqref{MR4D} and 
\eqref{Yukawa4D} can be rewritten in term of the matrix $F_{ij}$ as 
 \begin{eqnarray}
  M_{R,ij}         &=& F_{il}\kappa_{lm}F_{mj}v_{\rm B-L}, \label{MR4Dnew} \\
  \lambda_{i\alpha} &=& F_{in} \tilde{\lambda}_{n\alpha}. \label{Yukawa4Dnew}
 \end{eqnarray}
After the seesaw mechanism, the light neutrino mass matrix reads
 \begin{eqnarray}
  M_{\nu}=\lambda^{T} M_R^{-1} \lambda \langle \phi^0 \rangle^2 
  =\tilde{\lambda}^T\kappa^{-1}\tilde{\lambda}
   \frac{\langle \phi^0 \rangle^2}{v_{\rm B-L}}, \label{lightMajoranaMass}
 \end{eqnarray}
where the factor $f$ is canceled out in
the seesaw formula. By assigning the appropriate values for $m_i$, we can obtain
 one keV neutrino and two heavy ones naturally. For example, if we assume that 
the right-handed neutrino mass matrix is diagonal, we can realize a splitting 
mass spectrum of right-handed neutrinos, including e.g. both keV and very heavy 
mass scales, without fine-tuning as
 \begin{eqnarray}
  (M_{R,1},M_{R,2},M_{R,3})=(1\mbox{ keV},10^{11}\mbox{ GeV},10^{12}\mbox{ GeV}),
  \label{massspectrum}
 \end{eqnarray}
for $(m_1\ell,m_2\ell,m_3\ell)\simeq(24.2,3.64,2.26)$ where we take 
$M=5\times10^{17}$ GeV, $\ell^{-1}=10^{16}$ GeV, $v_{\rm B-L}=10^{15}$ GeV, 
$\kappa_{ii}=1$ as reference values, and $M_{R,i}$ is an effective right-handed 
neutrino mass. Therefore, the lightest right-handed neutrino with the mass of 
order keV scale can be a candidate for DM and heavy right-handed neutrinos with 
the masses of order $\mathcal{O}(10^{11-12})$ GeV can lead to the BAU via 
leptogenesis if the reheating temperature is larger than $\mathcal{O}(10^{11})$ 
GeV or $\mathcal{O}(10^{12})$ GeV unless the heavier guys are not extremely 
degenerated like in \eqref{massspectrum}. These are the essential points of the 
split seesaw mechanism.

\section{{\boldmath $A_4$} Flavor Models in Split Seesaw Mechanism}

An introduction of non-Abelian discrete flavor symmetry might be still one of 
fascinating approaches to explain the currently observed patterns of neutrino 
mixing angles rather than an approach that the current data are solely explained
 by determining relevant Yukawa couplings without symmetry and/or dynamics. 
Therefore, we attempt to embed such the symmetry into the split seesaw 
mechanism. The $A_4$ flavor symmetry is well motivated for the three generations
 of the SM fermions, as it is the smallest non-Abelian discrete group which has 
the triplet representation. 

\subsection{Embedding {\boldmath$A_4$} Flavor Models into Split Seesaw 
Mechanism}

A large number of flavor models based on the $A_4$ symmetry are widely discussed
 in literatures. Then an useful classification for $A_4$ flavor models has been 
recently presented in~\cite{Barry:2010zk}. We discuss the embedding $A_4$ flavor
 models into split seesaw mechanism with this Barry-Rodejohann (BR) 
classification for $A_4$ models, and show a simplified BR classification in 
Tab.~\ref{tab1}.\footnote{See~\cite{Barry:2010zk} for the references of 
corresponding flavor models.}
\begin{table}
\begin{center}
\begin{tabular}{cccc}
\hline
Type & $L_\alpha$                           & $\overline{E}_\alpha$               & $\overline{\Psi}_{iR}$              \\
\hline
\hline
A    & $\Rep{3}$                          & $\Rep{1}$, $\Rep{1}'$, $\Rep{1}''$ & $\cdots$                           \\
\hline
B    & $\Rep{3}$                          & $\Rep{1}$, $\Rep{1}'$, $\Rep{1}''$ & $\Rep{3}$                          \\
\hline
C    & $\Rep{3}$                          & $\Rep{3}$                          & $\cdots$                           \\
\hline
D    & $\Rep{3}$                          & $\Rep{3}$                          & $\Rep{3}$                          \\
\hline
E    & $\Rep{3}$                          & $\Rep{3}$                          & $\Rep{1}$, $\Rep{1}'$, $\Rep{1}''$ \\
\hline
F    & $\Rep{1}$, $\Rep{1}'$, $\Rep{1}''$ & $\Rep{3}$                          & $\Rep{3}$                          \\
\hline
G    & $\Rep{3}$                          & $\Rep{1}$, $\Rep{1}'$, $\Rep{1}''$ & $\Rep{1}$, $\Rep{1}'$, $\Rep{1}''$ \\
\hline
H    & $\Rep{3}$                          & $\Rep{1}$, $\Rep{1}$, $\Rep{1}$    & $\cdots$                           \\
\hline
I    & $\Rep{3}$                          & $\Rep{1}$, $\Rep{1}$, $\Rep{1}$    & $\Rep{1}$, $\Rep{1}$, $\Rep{1}$    \\
\hline
J    & $\Rep{3}$                          & $\Rep{1}$, $\Rep{1}$, $\Rep{1}$    & $\Rep{3}$                          \\
\hline
\end{tabular}
\end{center}
\caption{Simplified Barry-Rodejohann (BR) classification for $A_4$ flavor 
models}
\label{tab1}
\end{table}
That is classified by differences of particle assignments under $A_4$ symmetry 
for the lepton doublets $L$, right-handed charged leptons $E$, and right-handed 
neutrinos $\Psi_R$ (the Higgs triplets $\Delta$ are included in the original BR 
classification).

We are taking note of the right-handed neutrino mass spectrum in
 the split seesaw mechanism. Therefore, we discuss flavor models of Type B, D, 
E, F, G, I, and J because the models of Type A, C, and H do not contain the 
right-handed neutrinos. Moreover, it is remarked that the models of Type B, D, 
F, and J, which have assigned the right-handed neutrinos into the triplet 
$\Psi_R=(\Psi_{1R},\Psi_{2R},\Psi_{3R})^T$, cannot work in the context of split 
seesaw mechanism because the bulk mass term are degenerated,
\begin{equation}
m\bar{\Psi}_R\Psi_R=m(\bar{\Psi}_{1R}\Psi_{1R}+\bar{\Psi}_{2R}\Psi_{2R}
                      +\bar{\Psi}_{3R} \Psi_{3R}).
\end{equation}
Such a degenerated bulk mass spectrum cannot lead to a splitting mass spectrum of
 right-handed neutrinos as shown in \eqref{massspectrum}. Therefore, we can 
conclude that the suitable $A_4$ model for the split seesaw mechanism is that 
the three right-handed neutrinos should not be in the same multiplet. This is 
one of key observations in this work.

The model of Type E has been proposed by Ma~\cite{Ma:2005mw} and numerical 
analyses in the model has been presented in~\cite{Lavoura:2006hb}. This model is
 a simple extension of the SM to a model with $A_4$ flavor symmetry and three 
right-handed neutrinos represented by singlets under the symmetry. Therefore, we
 will consider the model further in the next subsection. The model of Type G has
 been proposed in~\cite{Frampton:2008ci}. The work has pointed out that possible
 mass spectrum of active neutrinos under the flavor charge assignment are 
$m_2\gg m_1=m_3=0$ or $m_3\gg m_1=m_2$. It is seen from the current neutrino 
experiments that they are ruled out. Therefore embedding the model of Type G 
into the split seesaw mechanism is not attractive. Finally, the model of Type I 
\cite{King:2006np} is the first complete supersymmetric model of flavor based on
 $A_4$ symmetry together with the $SU(4)_C\otimes SU(2)_L\otimes SU(2)_R$ 
Pati-Salam gauge symmetry. According to our simple discernment, the model can 
pass the above {\it degeneracy problem}. Since the model is supersymmetric 
version, which means that the model contains the lightest superparticle as a 
candidate for DM, embedding the mode of Type E proposed by Ma into the split 
seesaw mechanism is more attractive for our considerations. Therefore, we will 
consider a simple extension of the SM, which is similar to the model of type E, 
in the next subsection.

\subsection{Basic Model}

In this subsection, we consider an $A_4$ model with right-handed neutrinos 
assigned to singlets. The particle contents and charge assignment of the model 
discussed here are similar to the ones proposed by Ma in \cite{Ma:2005mw} 
except that in addition to $Z_2^{\rm aux}$ in Ma model, we add the auxiliary 
symmetry $Z_5^{\rm aux}$.\footnote{The auxiliary symmetry $Z_5^{\rm aux}$ plays
 an essential role in constraining the terms in the flavon scalar potential 
which will be shown in the following subsection.} The particle contents and 
charge assignment of the model are given in Tab.~\ref{tab:A4model}. 
\begin{table}
\begin{center}
\begin{tabular}{c||c|c|c|c|c||c||c|c|c|c}
\hline
Field         & $L_{\alpha}$ & $\overline{E}_{\alpha}$ & $\overline{\Psi}_{1R}$ & $\overline{\Psi}_{2R}$ & $\overline{\Psi}_{3R}$ & $\phi$    & $\varphi_{\nu,t}$    & $\varphi_{1l}$       & $\varphi_{2l}$     & $\varphi_{3l}$ \\
\hline
$A_4$         & $\Rep{3}$  & $\Rep{3}$             & $\Rep{1}$             & $\Rep{1}'$           & $\Rep{1}''$            & $\Rep{1}$ & $\Rep{3}$          & $\Rep{1}$           & $\Rep{1}''$        & $\Rep{1}'$ \\
\hline
$Z_2^{\rm aux}$ & $+$        & $-$    & $+$    & $+$    & $+$     & $+$       &  $+$ & $-$ & $-$ & $-$ \\
\hline
$Z_5^{\rm aux}$ & $\tilde{\omega}^3$        & $\tilde{\omega}^4$    & $1$    & $1$    & $1$     & $1$       &  $\tilde{\omega}^2$ & $\tilde{\omega}^3$ & $\tilde{\omega}^3$ & $\tilde{\omega}^3$ \\
\hline
\end{tabular}
\end{center}
\begin{center}
\caption{\label{tab:A4model} The particle contents and charge assignment of the 
model: $\varphi_{\nu,t}$ and $\varphi_{il}$ are flavons, which are assigned to the
 triplet and singlets, respectively. Note that $\tilde{\omega}\equiv 
e^{2\pi i/5}$.}
\end{center}
\end{table}

Now, let us first discuss the neutrino sector, the action is given in 
\eqref{Sneutrino}. The $\kappa$ and $\hat{\lambda}$ at the brane are given by
 \begin{equation}
  \kappa=\begin{pmatrix} 
          a & 0 & 0 \\ 
          0 & 0 & b \\ 
          0 & b & 0 
  \end{pmatrix},~~~  
  \tilde{\lambda}=\frac{1}{\Lambda}
                  \begin{pmatrix} 
                   y_1^{\nu} & 0       & 0       \\ 
                   0       & y_2^{\nu} & 0       \\ 
                   0       & 0       & y_3^{\nu} 
                  \end{pmatrix} 
                  \begin{pmatrix} 
                   1 & 1        & 1        \\ 
                   1 & \omega   & \omega^2 \\ 
                   1 & \omega^2 & \omega 
                  \end{pmatrix} 
                  \begin{pmatrix}
                    u_1 & 0  & 0  \\ 
                    0  & u_2 & 0  \\ 
                    0  & 0  & u_3 
                  \end{pmatrix}, \label{kappaAndlambda}
 \end{equation}
and the bulk mass $m$ becomes diagonal, $m=\mbox{Diag}\{m_1,m_2,m_3\}$, under 
the above charge assignment, where $\Lambda$ is a cut-off of the theory. The 
coupling $\tilde{\lambda}_{i \alpha}$ arise after the flavon $\varphi_{\nu,t}$ 
obtaining VEV as $\langle \varphi_{\nu,t} \rangle=(u_1,u_2,u_3)^T$,
 \begin{equation}
  \left[y_1^{\nu}\bar{\Psi}^{(0)}_{1R}\frac{(L \varphi_{\nu})}{\Lambda}
        +y_2^{\nu}\bar{\Psi}^{(0)}_{2R}\frac{(L \varphi_{\nu})''}{\Lambda} 
        +y_3^{\nu} \bar{\Psi}^{(0)}_{3R}\frac{(L \varphi_{\nu})'}{\Lambda}
  \right]\phi\rightarrow \tilde{\lambda}_{i\alpha}\bar{\Psi}^{(0)}_{iR}L_\alpha\phi.  
 \end{equation}
The neutrino Dirac mass matrix is given by
 \begin{eqnarray}
  M_D=\lambda\langle\phi^0\rangle=F\tilde{\lambda}\langle\phi^0\rangle
     =\frac{1}{\Lambda}F\begin{pmatrix} y_1^{\nu} & 0       & 0 \\ 
                                        0       & y_2^{\nu} & 0 \\ 
                                        0       & 0       & y_3^{\nu} 
                        \end{pmatrix} 
                        \begin{pmatrix} 1 & 1        & 1       \\ 
                                        1 & \omega   & \omega^2 \\ 
                                        1 & \omega^2 & \omega 
                        \end{pmatrix} 
                        \begin{pmatrix} u_1 & 0  & 0 \\ 
                                        0  & u_2 & 0 \\ 
                                        0  & 0  & u_3 
                        \end{pmatrix} v_{ew},
  \label{basic-D}
 \end{eqnarray}
where $\langle \phi^0 \rangle = v_{ew}$. The mass matrix of right-handed 
neutrinos in this model reads
 \begin{eqnarray}
  M_R=F\kappa F=\begin{pmatrix} 
                 a f_1^2 & 0        & 0 \\ 
                 0      & 0        & b f_2 f_3\\ 
                 0      & b f_2 f_3 & 0 
                \end{pmatrix}v_{B-L},
 \end{eqnarray}
which leads to the mass eigenvalues as
 \begin{eqnarray}
  M_{R,1}=a f_1^2 v_{B-L},~~~
  M_{R,2}=M_{R,3}=b f_2 f_3 v_{{\rm B-L}}.
  \label{spectrum}
 \end{eqnarray}
Inserting \eqref{kappaAndlambda} into \eqref{lightMajoranaMass}, the mass matrix
 of light neutrinos is given by 
 \begin{eqnarray}
  M_{\nu} =  \lambda^{T}\kappa^{-1}\lambda
            \frac{\langle\phi^0\rangle^2}{v_{{\rm B-L}}} 
         = \begin{pmatrix} 
            \frac{u_1}{\Lambda} & 0                  & 0 \\ 
            0                  & \frac{u_2}{\Lambda} & 0 \\ 
            0                  & 0                  & \frac{u_3}{\Lambda} 
           \end{pmatrix} 
           \begin{pmatrix} 
            x & y & y \\ 
            y & x & y \\ 
            y & y & x 
           \end{pmatrix} 
           \begin{pmatrix} 
            \frac{u_1}{\Lambda} & 0                  & 0 \\ 
            0                  & \frac{u_2}{\Lambda} & 0 \\ 
            0                  & 0                  & \frac{u_3}{\Lambda} 
           \end{pmatrix}\frac{v_{ew}^2}{v_{\rm B-L}}, \label{mnu}
 \end{eqnarray}
where 
 \begin{equation}
  x\equiv\frac{(y_1^\nu)^2}{a}+\frac{2y_2^{\nu}y_3^{\nu}}{b},~~~ 
  y\equiv\frac{(y_1^\nu)^2}{a}-\frac{y_2^{\nu}y_3^{\nu}}{b}. \label{xy}
\end{equation}

As discussed in \cite{Ma:2005mw}, the limit where $u_1=u_2=u_3=u$, the $A_4$ 
symmetry is broken to $Z_3$ in the neutrino sector and the eigenvalues of 
$M_{\nu}$ are 
 \begin{eqnarray}
  (x+2y)\frac{v_{ew}^2}{v_{\rm B-L}}\left(\frac{u}{\Lambda}\right)^2,~~~ 
  (x-y)\frac{v_{ew}^2}{v_{\rm B-L}}\left(\frac{u}{\Lambda}\right)^2,~~~
  (x-y)\frac{v_{ew}^2}{v_{\rm B-L}}\left(\frac{u}{\Lambda}\right)^2.
 \end{eqnarray} 
The neutrino mass matrix can be diagonalized by the tri-bimaximal mixing matrix.
 However, this case is not realistic since the first eigenvalue corresponds to 
the eigenstate $\frac{1}{\sqrt{3}}(1,1,1)$. Therefore, we have to consider the 
other case, that is, $u_1 \neq u_2 =u_3=u$. In this case, the mass matrix still 
leads to $\theta_{13}=0$ and $\theta_{23}=\pi/4$, while $\theta_{12}$ is determined
 by the parameters in the neutrino mass matrix. Using the reparametrization as
$u_1=ru$ and $u_2=u_3=u$, we can rewrite \eqref{mnu} as
 \begin{eqnarray}
  M_{\nu}=
   \begin{pmatrix} 
    r^2x & ry & ry \\ 
    ry   & x  & y  \\ 
    ry   & y  & x 
   \end{pmatrix}m_\nu~~~\mbox{ with }~~~ 
  m_\nu\equiv\frac{v_{ew}^2}{v_{\rm B-L}} \left(\frac{u}{\Lambda}\right)^2.
  \label{ModelTypeICon}
 \end{eqnarray} 
This mass matrix predicts $\theta_{13}=0$ and $\theta_{23}=\pi/4$, and the other 
observables are described by model parameters as follows  
 \begin{eqnarray}
  \Delta m_{21}^2 
   &\equiv& |m_2|^2-|m_1|^2=
            (x+r^2 x +y)\delta_{\nu}m_\nu^2, \\
  \Delta m_{31}^2 
   &\equiv& |m_3|^2-|m_1|^2=
   \left[(x-y)^2-\frac{[(1+r^2)x+y-\delta_\nu]^2}{4}\right] m_\nu^2, \\
   \tan^2\theta_{12} &=& \frac{[(1-r^2)x+y-\delta_{\nu}]^2}{8r^2y^2}, 
 \end{eqnarray}
where 
 \begin{eqnarray}
  \delta_{\nu}\equiv\sqrt{(1-r^2)^2x^2+2(1-r^2)xy+(1+8r^2)y^2},
 \end{eqnarray}
with real $r,x,$ and $y$. Now, let us fit the experimental data of neutrino 
oscillation, $\Delta m_{21}^2,$ $\Delta m_{31}^2$, and $\theta_{12}$ in this model.
 The best fit values~\cite{Schwetz:2011qt} for the normal mass hierarchy are
 \begin{eqnarray}
  \Delta m_{21}^2=7.64\times10^{-5}\mbox{ eV}^2,~~~
  \Delta m_{31}^2=2.45\times10^{-3}\mbox{ eV}^2,~~~
  \sin^2\theta_{12}=0.312. \label{exp}
 \end{eqnarray}
For example, the following values can reproduce the experimental 
values:\footnote{The model can also fit the inverted mass hierarchy (IH) by 
taking complex values of $r$, $x$, and $y$.} 
 \begin{eqnarray}
  (xm_\nu,ym_\nu,r)
  =(-2.23\times10^{-2}\mbox{ eV},~2.76\times10^{-2}\mbox{ eV},~\pm0.204).  
  \label{fit}
 \end{eqnarray}
Note that the model can also fit with the tri-bimaximal mixing, 
$\sin^2\theta_{12}=1/3$. For the charged lepton sector, the action is described 
by
 \begin{equation}
  S_l=\int d^4xdy\delta(y)
      \left((\epsilon \phi^*)
            \left[y_1^e(\overline{E}L)\frac{\varphi_{1l}}{\Lambda} 
                  +y_2^e(\overline{E}L)'\frac{\varphi_{2l}}{\Lambda} 
                  +y_3^e(\overline{E} L)''\frac{\varphi_{3l}}{\Lambda}
            \right]        
      \right).
 \end{equation}
After the flavons obtain their VEVs, it leads to a diagonal charged lepton mass 
matrix. We can also reproduce the charged lepton masses by taking appropriate 
values of $y_i^e\langle \varphi_{il} \rangle/\Lambda$.

Next, let us calculate the left-right mixing in this model, which is an 
important quantity in the sterile neutrino DM scenario. This scenario is a 
decaying DM candidate. To be DM, the lifetime of the lightest right-handed 
neutrino should be greater than the age of the Universe. The lightest 
right-handed neutrino can radiatively decay into a photon and an active neutrino
 through the left-right mixing. This radiative decay produces a narrow line in 
the diffuse X-ray background, which gives a restriction on the magnitude of this
 mixing angle~\cite{Boyarsky:2009ix}. The relevant left-right mixing angle is 
given by
 \begin{eqnarray}
  \theta^2 = \frac{\sum_{\alpha}|\lambda_{1 \alpha}|^2}{M_{R,1}^2}
               \langle\phi_0\rangle^2
           = (2+r^2)\frac{|y_1^{\nu}|^2}{a}\frac{m_\nu}{m_s} 
            <  1.8\times10^{-5}\left(\frac{1~{\rm keV}}{m_s}\right)^5,
\end{eqnarray}
where $m_s = M_{R,1}$. If we use $m_s=5~{\rm keV}$, $m_\nu=1.47\times10^{-3}$ eV, 
and $r=0.204$, we obtain an upper bound on model parameters as 
$|y_1^\nu|^2/a<9.63\times10^{-3}$. Note that the value of $m_\nu=1.47 \times10^{-3}$
 eV is obtained from $v_{ew}=174$ GeV, $v_{B-L}=10^{15}$ GeV, and 
$u/\Lambda\simeq0.22$. Using \eqref{xy}, \eqref{fit} and the value of $m_\nu$, 
we obtain $|y_1^{\nu}|^2/a=(x+2 y)/2\simeq11.3$, which is much larger than the 
above cosmological bound on a sterile neutrino DM scenario. The upper bound 
obtained here implies that the coupling to the keV sterile neutrino is too tiny 
and the only two heavy neutrinos can be taken into an account for the neutrino 
mass. Therefore, this $A_4$ model cannot realize a keV sterile neutrino DM 
scenario via the split seesaw mechanism though it can reproduce realistic mixing
 angles and two mass squared differences of active neutrinos. In the next 
subsection, we extend this $A_4$ flavor model to achieve the sterile neutrino DM
 scenario from split seesaw mechanism. 

\subsection{Extension of the Model}

In order to realize the keV sterile neutrino DM scenario in the split seesaw 
mechanism while conserving the realistic neutrino oscillation data, we extend 
the $A_4$ model by including the $SU(2)_L$ triplet Higgs and two gauge singlet 
flavons. The triplet Higgs and one of flavons are assigned to singlet under 
$A_4$ and the other flavon is a triplet. The extended particle contents is 
given in Tab.~\ref{tab3}.
\begin{table}
\begin{center}
\begin{tabular}{c||c|c||c}
\hline
Field        &  $\Delta_L$        & $\varphi_{\nu,s}$     & $\tilde{\varphi}_{\nu,t}$ \\
\hline
$A_4$        & $\Rep{1}$          & $\Rep{1}$           & \Rep{3}                 \\
\hline
$Z_2^{\rm aux}$ & $+$ &  $+$ & $+$                     \\
\hline
$Z_5^{\rm aux}$ & $\tilde{\omega}^2$ &  $\tilde{\omega}^2$ & $1$                     \\
\hline
\end{tabular}
\end{center}
\caption{Additional particle contents}
\label{tab3}
\end{table}
Note that the triplet flavon $\tilde{\varphi}_{\nu,t}$ will not contribute to the
 neutrino mass and mixing directly, however, it is crucial to obtain a VEV 
alignment for the another triplet flavon $\varphi_{\nu,t}$, which will be 
discussed in the next subsection.

The neutrino mass in this model is obtained from the type I+II seesaw mechanism.
 The neutrino sector with the contribution of type II seesaw 
mechanism~\cite{Lazarides:1980nt} is described by
 \begin{eqnarray}
  S_{\nu}=S_{\nu}^I+S_{\nu}^{II},
 \end{eqnarray}
where the type I contribution, $S_{\nu}^I$, is given in \eqref{Sneutrino} and the
 type II contribution is given by
 \begin{equation}
  S_{\nu}^{II}
   =\int d^4xdy\delta(y)\left(\tilde{y}^{\nu}_1\frac{\varphi_{\nu,t}}{\Lambda}
                              +\tilde{y}^{\nu}_2\frac{\varphi_{\nu,s}}{\Lambda}
                        \right)L^T\Delta_LL.   
\end{equation}
As discussed previously, the left-right mixing angle puts the upper bound on the
 coupling of the left-handed fields to the keV sterile neutrino, 
$|y_1^{\nu}|^2/a<9.63\times10^{-3}$, which is too tiny to be account for the 
neutrino mass scales. Keeping this constraint in mind, we obtain 
$x\simeq2y_2^{\nu}y_3^{\nu}/b$ and $y\simeq-y_2^{\nu}y_3^{\nu}/b$, resulting in 
$x\simeq-2y$. With the constraint on the left-right mixing angle, the type I 
contribution alone cannot produce the realistic neutrino oscillation data in 
this $A_4$ flavor model, therefore, we introduce the type II contribution. For 
the sake of the discussion, we will assume that $r=0$, that is, the VEV 
alignment of $\langle \varphi_{\nu,t} \rangle$ is given by 
$\langle\varphi_{\nu,t}\rangle=(0,u,u)^T$. We will show in the next subsection 
that this VEV alignment can be achieved from the orbifolding. Using $x=-2y$ and 
 $r=0$ , the neutrino mass matrix from type I contribution~\eqref{ModelTypeICon}
 becomes
 \begin{equation}
  M_{\nu}^I\simeq\begin{pmatrix} 
                0 & 0   & 0   \\ 
                0 & -2y & y   \\ 
                0 & y   & -2y 
               \end{pmatrix}  
               m_{\nu}. \label{Model2TypeImassNew}
\end{equation}
After the triplet Higgs obtains a VEV, 
$v_\Delta\equiv\langle\Delta_L\rangle\simeq v_{ew}^2 \mu/M_{\Delta}^2$, with 
the $B-L$ breaking scale $\mu \sim v_{\rm B-L}$ and the heavy triplet mass 
$M_\Delta$, and the flavor symmetry is broken by VEV of flavons, the neutrino 
mass from the type II seesaw mechanism is given by 
 \begin{eqnarray}
  M_\nu^{II}=\begin{pmatrix} 
            s & t & t \\ 
            t & s & 0 \\ 
            t & 0& s 
           \end{pmatrix} 
           m_\nu', \label{MII}
 \end{eqnarray}
where 
 \begin{eqnarray}
  s\equiv\tilde{y}^{\nu}_2\frac{\langle\varphi_{\nu,s}\rangle/\Lambda}{u/\Lambda},
  ~~~ 
  t\equiv\tilde{y}^{\nu}_1,~~~m_\nu'\equiv v_\Delta \frac{u}{\Lambda}.
  \label{parameters}
 \end{eqnarray}
Assuming that $m_\nu'\simeq m_\nu$, the neutrino mass matrix reads
 \begin{eqnarray}
  M_\nu=M_\nu^{I}+M_\nu^{II}
      \simeq\begin{pmatrix}  
             s & t    & t    \\ 
             t & s-2y & y    \\ 
             t & y    & s-2y 
            \end{pmatrix}m_\nu.
 \end{eqnarray}
This mass matrix has the $\mu-\tau$ symmetry resulting in $\theta_{13}=0$ and 
$\theta_{23}=\pi/4$ at the leading order.\footnote{The latest 
T2K results have suggested $0.03(0.04)<\sin^22\theta_{13}<0.28(0.34)$ at 90\% 
C.L. for NH (IH) the vanishing Dirac CP phase \cite{Abe:2011sj}. Then several 
theoretical discussions for the results have been presented \cite{He:2011kn}. 
Our model can give a non-vanishing contribution due to the next-leading order 
corrections. On the other hand, the recent MINOS results can be still 
consistent with the vanishing $\theta_{13}$ \cite{minos}.} The other 
observables can be calculated at the leading order as
 \begin{eqnarray}
  \Delta m_{21}^2    &\simeq& (2 s-y) \sqrt{8 t^2+y^2} m_\nu^2, \\
  \Delta m_{31}^2    &\simeq& \left[(s-3y)^2-\frac{\left(2s-y-\sqrt{8t^2+y^2}
                                                 \right)^2}{4}\right]m_\nu^2, \\
  \tan^2 \theta_{12} &\simeq& \frac{\left(y+\sqrt{8 t^2+y^2}\right)^2}{8t^2},
 \end{eqnarray}
with real $s$, $t$, and $y$. Now, we can fit the experimental values 
\eqref{exp} for the normal mass hierarchy as
 \begin{eqnarray}
  (s,t,y)\simeq(-1.09,-2.31,\pm2.06)\times10^{-2}\mbox{ eV}/m_\nu.
  \label{valuesII}
 \end{eqnarray}
The model can also fit the tri-bimaximal mixing angle with real $s$, $t$, and 
$y$, and the inverted mass hierarchy with complex values of the parameters. 
This $A_4$ model can reproduce realistic values of mixing angles and two mass 
squared differences of active neutrinos and be allowed by cosmological 
constraint on the left-right mixing angle in the context of keV sterile 
neutrino DM scenario from split seesaw mechanism.

It is well known that the keV sterile neutrino DM should not contribute to the 
neutrino mass via the type I seesaw mechanism to satisfy the X-ray bound. 
However, the $A_4$ flavor symmetry discussed in the previous subsection 
constrains on the Yukawa structure to be a specific form as given in 
\eqref{basic-D}. This means that the basic $A_4$ model cannot satisfy the 
cosmological constraint of the keV sterile neutrino DM while reproducing the 
observables of active neutrinos (mass squared differences and mixing angles) as
 we shown. However, once we introduce a contribution from the type II seesaw 
mechanism to the Majorana mass matrix of light neutrinos \eqref{MII}, the 
situation is changed, that is, the type II contribution can make the active 
neutrino observables realistic while realizing the keV sterile neutrino DM 
scenario. This is because the type II contribution to the Majorana mass matrix 
of light neutrinos is irrelevant to the sterile neutrino sector. This kind of 
scenario would be applied to any models (even with a flavor symmetry) where one
 of the three right-handed neutrinos is DM of keV mass (e.g. $\nu$MSM).
 
\subsection{Flavor Symmetry Breaking from Orbifolding} \label{FlavorOrbifold}

In this subsection, we show how the desired flavon VEV alignment can be obtained
 in the extra-dimensional context. The general discussion how to obtain the 
vacuum alignment for the flavon by orbifolding has been discussed before 
in~\cite{A4breaking} for $A_4$ in a different $A_4$ basis (see 
\cite{Adulpravitchai:2010na} for $S_4$ case).\footnote{See 
\cite{Holthausen:2009qj} for an alternative mechanism to flavor symmetry 
breaking in extra-dimensional setup. The work has proposed a general way to 
break flavor symmetry by imposing non-vanishing Dirichlet boundary condition on 
the bulk flavon(s), which can also achieve an arbitral VEV alignment.}

Let us consider a flavon field in the bulk. The desired flavon VEV alignment can
 be obtained by properly choosing the orbifold parity $P'$ and projecting out 
zero modes. Of course, there is a constraint on the choices of the orbifold 
parity, namely, it has to be chosen from elements of order two of $A_4$ group, 
otherwise, it does not preserve the product rule of the group. The elements of 
order two of $A_4$ group are as follows: $S$, $T^2 S T$, and $T S T^2$. To 
obtain the desired vacuum alignment, we choose the triplet flavon to transform 
under the orbifold parity $Z_2'$ as $P'=S$. 

We consider a bulk triplet flavon, 
$\varphi_\nu(y,x)=((\varphi_\nu)_1,(\varphi_\nu)_2,(\varphi_\nu)_3)^T$, transformed 
under the orbifold parities as 
 \begin{eqnarray}
  Z_2:~\varphi_\nu\rightarrow P\varphi_\nu=\eta\varphi_\nu,~~~
  Z_2':~\varphi_\nu\rightarrow P'\varphi_\nu=\eta'\varphi_\nu,
 \end{eqnarray}
where 
 \begin{eqnarray}
  P=1,~~~
  P'=S=\begin{pmatrix} 
        1 & 0  & 0  \\ 
        0 & -1 & 0  \\ 
        0 & 0  & -1 
       \end{pmatrix}.
\end{eqnarray}
Note that the flavor symmetry with this assignment is broken at the hidden 
($y=\ell$) brane, not at the SM ($y=0$) brane. However, the flavor symmetry is 
also broken at the SM brane in a low energy regime since non-zero modes in the 
triplet whose mass is the compactification scale are absent. The VEV alignment 
of flavon depends on the parity $\eta'$. When the parities $\eta=+1$ and 
$\eta'=+1$ are given for $\tilde{\varphi}_{\nu,t}$, the zero modes become 
$\tilde{\varphi}_{\nu,t}=((\tilde{\varphi}_{\nu,t})_1,0,0)^T$. When the parities 
$\eta=+1$ and $\eta'=-1$ are for $\varphi_{\nu,t}$, the zero modes read 
$\varphi_{\nu,t}=(0,(\varphi_{\nu,t})_2,(\varphi_{\nu,t})_3)^T$. In general, 
$(\varphi_{\nu,t})_2$ does not need to equal to $(\varphi_{\nu,t})_3$. But we show 
that $(\varphi_{\nu,t})_2 = \pm (\varphi_{\nu,t})_3$ can be obtained from the 
minimization of the 4D flavon scalar potential in the following discussion.

At low energy, the effective (4D) flavon scalar potential is described by the 
zero modes of the flavon,
\begin{eqnarray}
  V_{f}^{\rm (4D)} &=& \frac{m_1^2}{2}(\tilde{\varphi}_{\nu,t})_1^2 
                    +m_2^2 (|(\varphi_{\nu,t})_2|^2+|(\varphi_{\nu,t})_3|^2) 
                    +[\mu_{123}((\tilde{\varphi}_{\nu,t})_1(\varphi_{\nu,t})_2^*
                      (\varphi_{\nu,t})_3+{\rm h.c.}] \nonumber \\
                   & & +m_0^2|(\varphi_{\nu,s})|^2 
                       +\tilde{m}_1^2|(\varphi_{1l})|^2
                       +\tilde{m}_2^2 |(\varphi_{2l})|^2 
                       +\tilde{m}_3^2 |(\varphi_{3l})|^2
                       +\sum_{i,j}\lambda_{ij}X_i X_j,
 \end{eqnarray}
where $X=((\tilde{\varphi}_{\nu,t})_1^2,(|(\varphi_{\nu,t})_2|^2+|(\varphi_{\nu,t})_3|^2),|(\varphi_{\nu,s})|^2,|(\varphi_{1l})|^2,|(\varphi_{2l})|^2,|(\varphi_{3l})|^2)^T$ and $\lambda_{ij}$ are dimensionless couplings. Note that the 
normalization constant of the wave function and the cut-off are already 
included into the flavon scalar potential parameters and we assume that all 
singlet flavons are the bulk fields.

The VEV alignment can be obtained by the minimization of the flavon scalar 
potential, $\partial V_{f}^{\rm (4D)}/(\partial \varphi_i)=0$, where $\varphi_i$ 
are flavons in the potential. We can see that the condition 
$(\varphi_{\nu,t})_2 = \pm (\varphi_{\nu,t})_3$ can be achieved if $\mu_{123}$ and 
$(\tilde{\varphi}_{\nu,t})_1$ are non-zero. This is the reason we introduce 
$\tilde{\varphi}_{\nu,t}$ even it does not give the contributions to the neutrino
 mass and mixing at the leading order. 

\subsection{Leptogenesis}

At the end of this paper, we comment on the realization of BAU via leptogenesis 
in our $A_4$ model.

\begin{figure}
\begin{center}
\includegraphics[scale = 1.41]{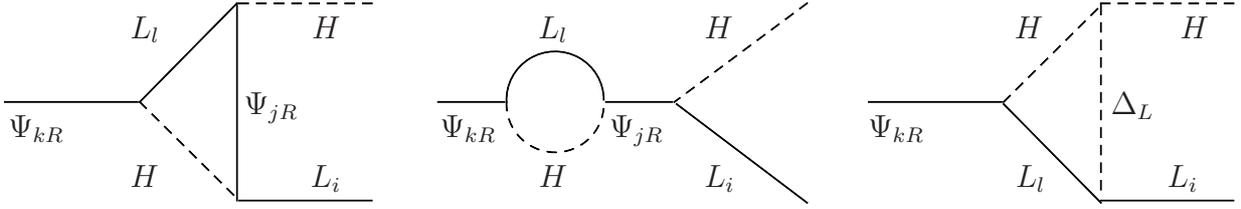}
\end{center}\vspace{-3.2cm}

\hspace{1.7cm}$L_l$\hspace{2cm}$H$\hspace{2.6cm}$L_l$\hspace{1.8cm}$H$\hspace{3.7cm}$H$\hspace{1.8cm}$H$\vspace{0.5cm}

\hspace{3.2cm}$\Psi_{jR}$\hspace{10.7cm}$\Delta_L$\vspace{-0.2cm}

\hspace{0.1cm}$\Psi_{kR}$\hspace{4.95cm}$\Psi_{kR}$\hspace{1.5cm}$\Psi_{jR}$\hspace{2.7cm}$\Psi_{kR}$\vspace{1.5mm}

\hspace{1.7cm}$H$\hspace{2cm}$L_i$\hspace{2.6cm}$H$\hspace{1.8cm}$L_i$\hspace{3.7cm}$L_l$\hspace{1.6cm}$L_i$
\caption{One-loop diagrams of the right-handed neutrino decay generating the CP 
asymmetry.}
\label{fig1}
\end{figure}

In our model, the two heavier right-handed neutrinos are not the only source for the BAU via leptogenesis, but we also have the contribution from
 the triplet Higgs~\cite{O'Donnell:1993am}. Leptogenesis can be realized 
either by the decay of the heavier right-handed neutrinos or the triplet Higgs. 
The CP asymmetry in the decay of those right-handed neutrinos to the leptons and
 SM Higgs is given by
 \begin{eqnarray}
  \epsilon_{\Psi_{kR}}
   =\sum_\alpha\frac{\Gamma(\Psi_{kR}\rightarrow L_i+H)
                    -\Gamma(\Psi_{kR}\rightarrow \bar{L}_i+H^\dagger)}
                  {\Gamma(\Psi_{kR}\rightarrow L_i+H)
                   +\Gamma(\Psi_{kR}\rightarrow \bar{L}_i+H^\dagger)},
 \end{eqnarray}
where $k=2$ and $3$ because the decay of heavier neutrinos is CP asymmetric 
in the split seesaw mechanism. As usual, this asymmetry is due to the 
interference of the tree level diagram 
and the three one-loop diagrams shown in 
Fig.~\ref{fig1}.
The left two diagrams are the usual 
vertex and self-energy contributions, 
respectively, and give
 \begin{eqnarray}
  \epsilon_{\Psi_{kR}}^I
   =\frac{1}{8\pi}\sum_{j\neq k}
    \frac{\mbox{Im}[(\lambda\lambda^\dagger)_{kj}^2]}
         {\sum_i|\lambda_{ki}|^2}\sqrt{x_j}
    \left[1-(1+x_j)\log\left(\frac{x_j+1}{x_j}\right)+\frac{1}{1-x_j}\right],
 \end{eqnarray}
where $x_j\equiv M_{R,j}^2/M_{R,k}^2$.\footnote{The mass spectrum of the 
right-handed neutrinos in our model is given in \eqref{spectrum}, which is degenerate at the leading order. However, the next leading correction, whose 
magnitude is typically $\mathcal{O}((u/\Lambda)^2 M_{R,3})$, weaken this 
degeneracy to be $|M_{R,3}-M_{R,2}| \gg \Gamma_{\Psi_{3R}}$. Therefore, the hierarchical 
formulae for the computation of 
the CP asymmetry reviewed in this section are valid 
for our model.} The right diagram in Fig.~\ref{fig1} is an additional 
contribution to the CP asymmetry in our model, which is given by 
\cite{O'Donnell:1993am,Antusch:2004xy} as
 \begin{eqnarray}
  \epsilon_{\Psi_{kR}}^{II}
   =\frac{3}{8\pi}\frac{M_{R,k}v_\Delta}{v_{ew}^2}\sum_{j,l}
    \frac{\mbox{Im}[\lambda^\ast_{kj}\lambda^\ast_{kl}Y_\Delta^\ast{}_{jl}]}
         {\sum_i|\lambda_{ki}|^2}y_k\left[1-y_k\log\left(\frac{y_k+1}{y_k}
    \right)\right],
 \end{eqnarray}
where $Y_\Delta\equiv M_\nu^{II}/v_\Delta$ and $y_k\equiv M_\Delta^2/M_{R,k}^2$. Note 
that there is also a decay process to generate the asymmetry, which is the decay
 of the triplet Higgs to two leptons. However, it is not important for our case 
as commented below.

There are four limiting cases for leptogenesis in the type I$+$II seesaw 
scenario~\cite{O'Donnell:1993am}: [i] $M_{R,k}\ll M_\Delta$ with a dominant 
contribution of the right-handed neutrinos to the light neutrino masses, [ii] 
$M_{R,k}\ll M_\Delta$ with a dominant triplet contribution to the light neutrino 
masses, [iii] $M_{R,k}\gg M_\Delta$ with a dominant right-handed neutrino 
contribution to the neutrino masses, and [iv] $M_{R,k}\gg M_\Delta$ with dominance
 of the triplet contribution to the neutrino masses, where $M_{R,k}$ is the mass of 
the heavier right-handed neutrino. 
The limit [i] is almost equivalent to the ordinary 
leptogenesis scenario in the framework of type I seesaw mechanism, and thus this
case has been extensively studied in the literatures. In the limit [ii], a 
constraint on the neutrino mass does not induce a violation of the 
out-of-equilibrium condition, $\Gamma_{M_{R,k}}<H(T)|_{T=M_{R,k}}$, where $H(T)$ is 
the Hubble parameter and $T$ is the cosmic temperature, since the couplings 
responsible for the neutrino masses are not responsible for the tree level 
decay. As a result, there is no upper bound on the neutrino masses as in the 
usual type I leptogenesis, which is an interesting feature of the limit [ii]. The
 decay process of triplet Higgs to two leptons become important for the limits 
[iii] and [iv]. Our model naively corresponds to a mixed scenario of limits [i] 
and [ii], that is, $M_{R,k}\ll M_\Delta$ with comparable contributions of the 
right-handed neutrinos and triplet Higgs to the light neutrino mass because of 
\eqref{spectrum}, \eqref{ModelTypeICon}, \eqref{parameters}, \eqref{valuesII} 
and the assumption $m_\nu'\simeq m_\nu$. Because of the presence of additional 
contributions from the triplet Higgs decay to CP asymmetry, the bounds on the decay 
asymmetry and on $M_{R,2}$ become weaker than the ones obtained 
in~\cite{Antusch:2004xy}, which discussed the limit [i]. Therefore, of course, 
our model with contributions from type I and II seesaw mechanisms can also 
realize the mixed scenario of type I$+$II leptogenesis. Solving the exact 
Boltzmann equation and searching for all allowed regions of model parameters are
 out of our scope in this paper.

\section{Conclusion}

We have studied models based on $A_4$ flavor symmetry in the context of the 
split seesaw mechanism. The split seesaw mechanism, which has been proposed an 
extra-dimensional theory, can provide a natural way to realize a splitting mass 
spectrum of right-handed neutrinos. It leads to one keV sterile 
neutrino as a DM candidate favored by recent cosmological observations and two 
heavy right-handed neutrinos being responsible for leptogenesis to explain the 
observed BAU.

First, we have pointed out that most of known $A_4$ flavor models with three 
right-handed neutrinos being $A_4$ triplet suffer from a {\it degeneracy 
problem} for the bulk mass term, which disturbs the split mechanism for 
right-handed neutrino mass spectrum.

Next, we have constructed a new $A_4$ flavor model to work in the split seesaw 
mechanism. The model is an extention of the model proposed by Ma. The three 
right-handed neutrinos are assigned into $A_4$ (non-trivial) singlets, and the 
neutrino masses and mixing angles can be realized from both type I$+$II seesaw 
contributions in the model. The model predicts the $\mu-\tau$ symmetry in the 
neutrino mass matrix at the leading order, resulting in the vanishing 
$\theta_{13}$ and maximal $\theta_{23}$. The flavor symmetry $A_4$ is broken via 
the flavon vacuum alignment which can be obtained from the orbifold 
compactification. The BAU can also be realized in the mixed scenario of type 
I$+$II leptogenesis.

\subsection*{Acknowledgement}

We are grateful to Takehiko Asaka, Peihong Gu, Claudia Hagedorn, Andreas 
Hohenegger, Martin Holthausen, Alexander Kartavtsev, Manfred Lindner, Kristian 
McDonald, and Werner Rodejohann for fruitful discussions. This work has been 
supported by the DFG Sonderforschungsbereich Transregio 27 Neutrinos and beyond
 Weakly interacting particles in Physics, Astrophysics and Cosmology.

\appendix
\section*{Appendix}
\subsection*{Group Theory of {\boldmath $A_4$}}
We give a brief knowledge of group theory of $A_4$ in this Appendix.

The order of $A_4$ group is 12. It contains 12 group elements which can be 
obtained from the combinations of the two group generators, $S$ and $T$, which 
obey the generator relations,
 \begin{equation}
  S^2 = 1, \;\;\; T^3=1, \;\; \; \;{\rm and} \;\; (ST)^3=1. \nonumber
 \end{equation}
The group contains three singlets and one triplet as its irreducible 
representations. The singlets transform under the group generators as
 \begin{eqnarray}
  \begin{array}{lccl}
   \Rep{1}   & : &  S=1, & T=1,                  \\
   \Rep{1}'  & : &  S=1, & T=e^{i 2 \pi/3}=\omega,  \\
   \Rep{1}'' & : &  S=1, & T=e^{i 4 \pi/3}=\omega^2, 
   \end{array}
  \nonumber 
 \end{eqnarray}
and the triplet transforms as 
 \begin{eqnarray}
  \Rep{3}: &   S=\begin{pmatrix} 
                  1 & 0  & 0  \\ 
                  0 & -1 & 0  \\ 
                  0 & 0  & -1 
                 \end{pmatrix}, & 
               T=\begin{pmatrix} 
                  0 & 1 & 0 \\ 
                  0 & 0 & 1 \\ 
                  1 & 0 & 0 \end{pmatrix}. \nonumber 
 \end{eqnarray}
The products of singlets are given as $\Rep{1}' \times \Rep{1}'=\Rep{1}''$, 
$\Rep{1}' \times \Rep{1}''=\Rep{1}$, $\Rep{1}'' \times \Rep{1}''=\Rep{1}''$ 
etc.. When we represent the triplet as $\Rep{3}_a = (a_1,a_2,a_3)^T$ and  
$\Rep{3}_b = (b_1,b_2,b_3)^T$, the product of two triplets gives 
$\Rep{3}_a\times\Rep{3}_b=\Rep{1}+\Rep{1}'+\Rep{1}''+\Rep{3}+\Rep{3}$, where
 \begin{eqnarray}
  \begin{array}{lcl}
   \Rep{1}   & = & a_1b_1+a_2b_2+a_3b_3,                \\
   \Rep{1}'  & = & a_1b_1+\omega^2a_2b_2+\omega a_3b_3, \\
   \Rep{1}'' & = & a_1b_1+\omega a_2b_2+\omega^2a_3b_3, \\
   \Rep{3}   & = & (a_2 b_3, a_3 b_1, a_1 b_2)^T,       \\
   \Rep{3}   & = & (a_3 b_2, a_1 b_3, a_2 b_1)^T. 
  \end{array} \nonumber
 \end{eqnarray} 


\end{document}